\documentclass[prl,twocolumn,floats,aps,epsfig,nofootinbib,amssymb]{revtex4}
\usepackage{subfigure}
\usepackage[dvips]{graphicx}
\usepackage{amsmath}
\usepackage{bm}
\usepackage{times}
\usepackage{epsfig}
\begin{document}

\title{SPONTANEOUS R-PARITY BREAKING AND LEFT-RIGHT SYMMETRY}
\bigskip
\author{Pavel Fileviez P{\'e}rez}
\author{Sogee Spinner}
\address{
Department of Physics, University of Wisconsin, Madison, WI 53706, USA}
\date{\today}

\begin{abstract}
We propose a simple renormalizable left-right theory where R-parity
is spontaneously broken and neutrino masses are generated through
the Type I seesaw mechanism and R-parity violation. In this theory R-parity and
the gauge symmetry are broken by the sneutrino vacuum expectation
values and there is no Majoron problem. The $SU(2)_R$ and R-parity
violation scales are determined by the SUSY breaking scale making
the model very predictive. We discuss the spectrum and possible
tests of the theory through the neutralinos, charginos,
$Z^{'}$ and $W_R^{\pm}$ decays at the Large Hadron Collider.
\end{abstract}
\maketitle
\section{I. Introduction}
The existence of massive neutrinos, the unknown origin of parity violation in the
Standard Model (SM) and the hierarchy problem are some of the main motivations
for physics beyond the SM. In the context of the so-called left-right
symmetric theories~\cite{LR} one has the appealing possibility to understand the origin
of parity violation and its strong connection to the generation of neutrino masses.
The supersymmetric version of these theories can also
solve the hierarchy problem, as in the Minimal Supersymmetric Standard Model (MSSM).

Defining a minimal left-right model is a subtle issue since particle content depends on
the mechanism generating neutrino masses.  In the so-called minimal
left-right symmetric theory, neutrino masses are generated through the Type I~\cite{TypeI}
and Type II~\cite{TypeII} seesaw mechanisms.  Alternatively, it is possible to have a simple
theory~\cite{LR-TypeIII} where neutrino masses are generated through the
Type I~\cite{TypeI} and Type III~\cite{TypeIII} seesaw mechanisms.

Supersymmetric left-right particle content further depends on the status of R-parity,
an \textit{ad hoc} discrete symmetry imposed in the MSSM to forbid rapid proton decay.
Above the left-right scale, R-parity is automatically conserved due to local $U(1)_{B-L}$.
Two situations are possible for the low energy theory: automatic R-parity conservation
or spontaneous R-parity violation, which conserves baryon number and therefore does not
induce proton decay. The former was discussed in Ref.~\cite{LR-Models} where the
necessary Higgs sector was found to be involved but parity is spontaneously broken.
The latter can have a simpler Higgs sector as well as exciting collider predictions,
making the origin and impact of R-parity breaking in such models an important issue.

In this Letter we investigate this issue in detail and find that the Higgs sector can be
remarkably simplified. This simplification utilizes the right-handed sneutrino which has
both $B-L$ and $SU(2)_R$ quantum numbers. Once this field acquires a vacuum expectation
value (VEV), both $SU(2)_R \bigotimes U(1)_{B-L}$  and R-parity are spontaneously broken
with the relevant scales determined by the soft SUSY breaking scale. This leads to a
simple and predictive model, especially for decays of the neutralinos, charginos, $Z^{'}$ and $W_R$.
Neutrino masses are generated via the Type-I seesaw mechanism and R-parity. The Majoron
problem~\cite{Majoron} associated with spontaneous lepton number breaking is not present since the Majoron
becomes the longitudinal component of $Z^{'}$. Furthermore, all of this could be 
accomplished with the same Higgs sector of the MSSM making this the simplest left-right symmetric theory
without R-parity. In this theory the left-right discrete symmetry is broken only
by the soft terms in order to have a consistent mechanism for R-Parity violation
and avoid the domain wall problem.

This paper is organized as follows: in Section II we
discuss the theory, while in Section III we show the
properties of the spectrum, the R-parity violation
mechanism, and the mechanism generating neutrino masses
are discussed in great detail. The possible tests of the theory
are discussed in Section IV.
\section{II. Minimal SUSY Left-Right Theory and R-Parity}
Left-right symmetric theories are based on the gauge group
$SU(3)_C \bigotimes SU(2)_L \bigotimes SU(2)_R \bigotimes U(1)_{B-L}$.
Here $B$ and $L$ stand for baryon and lepton number, respectively.
In the supersymmetry (SUSY) case, the matter chiral supermultiplets for quarks and
leptons are given by

\begin{equation}
\hat{Q} = \left(
\begin{array} {c}
\hat{U} \\ \hat{D}
\end{array}
\right) \ \sim \ (2,1,1/3),
\
\hat{Q}^C = \left(
\begin{array} {c}
 \hat{U}^C \\ \hat{D}^C
\end{array}
\right) \ \sim \ (1,2,-1/3),
\end{equation}
\begin{equation}
\hat{L} = \left(
\begin{array} {c}
 \hat{N} \\ \hat{E}
\end{array}
\right) \ \sim \ (2,1,-1),
\ \text{and} \
\hat{L}^C = \left(
\begin{array} {c}
 \hat{N}^C \\ \hat{E}^C
\end{array}
\right) \ \sim \ (1,2,1),
\end{equation}
where $N^C$, the right-handed neutrino, is now required by the gauge group.
With this field content, the superpotential is:
\begin{eqnarray}
	\label{W}
	{\cal W} 	&	=	&
		 Y_q \ \hat{Q}^T \ i \sigma_2 \ \hat{\Phi}  \ i \sigma_2 \ \hat{Q}^C \ + \
		 Y_\nu^D \ \hat{L}^T \ i \sigma_2 \ \hat{\Phi} \ i \sigma_2 \ \hat{L}^C
	\nonumber
	\\
		& + & \frac{\mu}{2} \text{Tr} \left( \hat{\Phi}^T \ i \sigma_2 \ \hat{\Phi} \ i \sigma_2 \right),
\end{eqnarray}
where the bi-doublet Higgs is defined by
\begin{equation}
\hat{\Phi} = \left(
\begin{array} {cc}
 \hat{H}_D^0   &  \hat{H}^+_U \\
 \hat{H}_D^-  & \hat{H}^0_U
\end{array}
\right) \ \sim \ (2, 2, 0).
\end{equation}
Typically, in order to have consistent relationships between
the quark masses, an extra bi-doublet needs to be introduced
or one-loop gluino corrections to the quark
masses through trilinear soft breaking terms that are different between
the up and down quark sectors must be assumed~\cite{Mohapatra}. 
Here we see both possibilities as appealing. 

So far, it seems like extra superfields are needed to
break $SU(2)_R \times U(1)_{B-L}$, since the bi-doublet does not have a $B-L$ 
quantum number. As mentioned earlier, the choice for these
fields depends on whether or not the low energy theory should conserve R-parity (or M-Parity).
R-parity is defined as $R=(-1)^{3(B-L) + 2 S}=(-1)^{2S}M$, where
M is M-parity. As it is well known M-parity is $-1$
for any matter chiral superfield and $+1$ for any Higgs or vector
superfield. Therefore, R-parity conservation requires Higgs fields
with an even value of $B-L$.  Typically, this is achieved by introducing several
extra Higgs chiral superfields or higher-dimensional
operators as in~\cite{LR-Models}.

In the Letter, we wish to take advantage of the fact that
Eq.~(\ref{W}) already contains a scalar field with the correct quantum
numbers: the right-handed sneutrino. Once this field acquires a
VEV, it spontaneously breaks both the higher gauge symmetry as well as
R-parity and forces left-handed sneutrino, through mixing terms,
to acquire a VEV. Since lepton number is part of the gauge symmetry the 
Majoron (the Goldstone boson associated with spontaneous
breaking of lepton number) becomes the longitudinal component of the $Z^{'}$ and
does not pose a problem. Therefore, in this context one can have a simple and
consistent TeV scale theory for spontaneous $SU(2)_R \times U(1)_{B-L}$
and R-parity violation with the same Higgs sector as the MSSM.

The kinetic terms in the theory are given by
\begin{eqnarray}
	{\cal L}_{Kin} &=& \int d^4\theta
		\ \text{Tr} \left( \hat{\Phi}^\dagger \ e^{g_L \hat{V}_L} \ \hat{\Phi} \ e^{g_R \hat{V}_R} \right)
	\nonumber
	\\
		 & + & \int d^4 \theta \ \hat{L}^\dagger e^{g_L \hat{V}_L \ - \frac{1}{2} \ g_{BL} \hat{V}_{BL}} \hat{L} \nonumber \\
		 & + & \int d^4 \theta \ \hat{L^C}^\dagger e^{g_R \hat{V}_R^T \ + \frac{1}{2} \ g_{BL} \hat{V}_{BL}} \hat{L}^C,
\end{eqnarray}
where $\hat{V}_{L}$ and $\hat{V}_{R}$ are the vector superfields
for the gauge bosons in $SU(2)_L$ and $SU(2)_R$, respectively.
Here, we use $g_L$ and $g_R$ for the gauge couplings in left-right
sector.

In our notation the soft breaking terms are given by

\begin{eqnarray}
	V_{soft} &	=&
		M_{\tilde Q}^2 \ \tilde{Q}^\dagger \tilde{Q} \ + \  M_{\tilde Q^C}^2 \ \tilde{Q^C}^\dagger \tilde{Q^C} \ + \
		 M_{\tilde L}^2 \ \tilde{L}^\dagger \tilde{L}
	\nonumber
	\\
		& + & M_{\tilde L^C}^2 \ \tilde{L^C}^\dagger \tilde{L^C}  \ + \ M_{\Phi}^2 \ \text{Tr} \left( \Phi^\dagger \Phi \right)
	\nonumber
	\\
		 & + &
		 \left(
		 	\frac{1}{2} M_R \text{Tr} \ \tilde{W}_R \tilde{W}_R \ + \ \frac{1}{2} M_L \text{Tr} \ \tilde{W}_L \tilde{W}_L
		\right.
	\nonumber
	\\
		& &
		\left.
			+ \ \frac{1}{2} M_{BL} \tilde{B} \tilde{B} \ + \ A_q^1 \ \tilde{Q}^T \ i \sigma_2 \ \Phi \ i \sigma_2 \tilde{Q}^C \
		\right.
	\nonumber
\\
		& &
		\left.
			+ \ A_\nu^D \ \tilde{L}^T \ i \sigma_2 \ \Phi \ i \sigma_2 \ \tilde{L}^C \
		\right.
	\nonumber
	\\
		& &
		\left.
			+ \ B \ \frac{\mu}{2} \ Tr \left( \Phi^T i \sigma_2 \Phi i \sigma_2 \right) \ + \ h.c.
		\right)
\label{soft}
\end{eqnarray}
It is important to mention that under the discrete Left-Right Symmetry
one has the transformations: $\hat{Q} \leftrightarrow \hat{Q^C}^*$,
$\hat{L} \leftrightarrow \hat{L^C}^*$ and $\hat{\Phi} \leftrightarrow \hat{\Phi}^\dagger$.
In this case the Yukawa couplings $Y_q$ and $Y_\nu^D$ are hermitian. Notice that in general
there is no reason to assume the left-right discrete symmetry in the soft-breaking sector.
In the rest of the Letter we will assume that the Left-Right discrete symmetry
is only softly broken by the soft terms. In this case one can have a consistent
mechanism for R-parity violation and avoid the domain wall problem.
Notice that the breaking of the Left-Right symmetry is transmitted
only through loop effects to the gauge interactions relevant for $\beta$
and $\mu$ decays. Also we can add non-holomorfic soft terms 
$A_q^2 \ \tilde{Q}^T \ \Phi^* \ \tilde{Q}^C \ + \ \ A_\nu \ \tilde{L}^T \ \Phi^* \ \tilde{L}^C $ 
which could help us to correct the relation between the fermion masses at one-loop.  
\section{III. Spectrum and R-parity Violation}
In this theory the gauge boson masses are generated by
the vacuum expectation values (VEVS) of sneutrinos
($\langle \tilde{\nu_i} \rangle=v_L^i/\sqrt{2}$
and $\langle \tilde\nu^C_i \rangle=v_R^i/\sqrt{2}$) and the bi-doublet
($\left< H_U^0 \right> = v_u/\sqrt{2}$ and $\left< H_D^0 \right> = v_d/\sqrt{2}$).
The sneutrino VEVs also break R-parity and lepton number eliminating
the quantum numbers necessary to distinguish between the
lepton, Higgs and gaugino sectors. Therefore the physical
charginos and neutralinos, as well as the Higgses will be admixtures
of these three sectors. The properties
of the full spectrum will be discussed in more detail in a future
publication~\cite{Spinner}. It is important to mention that in this
context there is no Majoron problem since lepton number
is part of the gauge symmetry (the Majoron becomes the longitudinal component of $Z^{'}$).

The scalar potential in this theory is given by
\begin{eqnarray}
	V & = & V_{F} \ + \ V_D \ + \ V_{soft}^S,
\end{eqnarray}
where the relevant terms for $V_{soft}^S$ are given
in Eq.~(\ref{soft}). Once one generation of sneutrinos, $\tilde{\nu}$ and
$\tilde{\nu}^C$, and $\Phi$,  acquire a VEV, the potential reads
\begin{align}
	\notag
	\left<V_F \right>	&	=
		\frac{1}{4} \left(Y_\nu^D \right)^2
		\left(
			v_R^2 v_u^2 + v_R^2 v_L^2 + v_L^2 v_u^2
		\right)
		+ \frac{1}{2} \mu^2
		\left(
			v_u^2 + v_d^2
		\right)
	\\
	&
		+ \frac{1}{\sqrt{2}} Y_\nu^D \ \mu \ v_R v_L v_d
	\\
	\notag
	\left<V_D \right>	&	=
		\frac{1}{32}
		\left[
			g_R^2
			\left(
				v_R^2 + v_d^2 - v_u^2
			\right)^2
			+ g_L^2
			\left(
				 v_u^2 -v_d^2 - v_L^2
			\right)^2		
		\right.
	\\
	&
		+ \left.
			g_{BL}^2
			\left(
				v_R^2 - v_L^2
			\right)^2
		\right]
	\\
	\notag
	\left<V_{soft}^S \right>	&	=
		\frac{1}{2} M_{\tilde L}^2 v_L^2 + \frac{1}{2} M_{\tilde L^c}^2 v_R^2 + \frac{1}{2} M_{\Phi}^2 \left(v_u^2 + v_d^2 \right)
	\\
	&
		- \text{Re} \left( B \mu \right) \ v_u v_d  \nonumber
\\
&
- \frac{1}{2 \sqrt{2}}
\left(A_\nu^D + \left(A_\nu^D \right)^\dagger \right) v_R v_L v_u
\end{align}
and can be minimized in the usual way.
Illuminating results can be found for  the case $v_R \gg v_u, v_d \gg v_L$, (a reasonable assumption given the
phenomenologically necessary hierarchy between the left-and right-handed scales):
\begin{align}
	\label{MC.vR}
	v_R &	=
		\sqrt{\frac{- 8 M_{\tilde{L}^c}^2}{g_R^2 + g_{BL}^2}}
	\\
	\label{MC.vL}
	v_L &	=
		\frac{A_\nu^D \ v_R v_u}{\sqrt{2}
		\left(
			M_{\tilde L}^2 - \frac{1}{8} g_{BL}^2 v_R^2
		\right)
		}
	\\
	\label{MC.mu}
	\mu^2 &	=
		-\frac{1}{8} \left(g_R^2 + g_L^2 \right) \left(v_u^2 + v_d^2 \right) +
		\frac
		{
			M_{H_U}^2 \tan^2 \beta - M_{H_D}^2
		}
		{
			1 - \tan^2 \beta
		}
	\\
	\label{MC.Bmu}
	B\mu	&	=
		\frac{\sin{2 \beta}}{2}
		\left(
			2 \mu^2 + M_{H_U}^2 + M_{H_D}^2
		\right)
\end{align}
where Eq.~(\ref{MC.vR}) has the same form as the Standard Model minimization
condition and demonstrates the need for $M_{\tilde{L}^c}^2 < 0$, while Eq.~(\ref{MC.vL})
indicates that $A_\nu^D$ should be small, \textit{i.e.}
$A_\nu^D  \ll \frac{\sqrt{2}}{8} g_{BL}^2 \frac{v_R^2}{v_u}$
in order to
have $v_R \gg v_L$.  Equations~(\ref{MC.mu}) and ~(\ref{MC.Bmu}) are similar to their
MSSM counterparts with $M_{H_U}^2 \equiv M_\Phi^2 - \frac{1}{8} g_R^2 v_R^2$ and
$M_{H_D}^2 \equiv M_\Phi^2 + \frac{1}{8} g_R^2 v_R^2$.  Note that the negative
contribution to $m_{H_U}^2$ is conducive for electroweak symmetry breaking.
Also, even though $M_{\tilde{L}^c}^2$ is negative, a realistic spectrum still exists~\cite{Spinner}.
\subsection{III.A. Neutrino Masses}
Let us discuss how the neutrino masses are generated
in this context. Once the symmetry is broken the
neutralinos $\tilde{\chi}^0$ are defined as a linear
combination of $\tilde{B}$, $\tilde{W}_R^0$, $\tilde{H}_D^0$,
$\tilde{H}_U^0$ and $\tilde{W}_L^0$ and their mass matrix
is given by
\begin{equation}
{\cal M}_{\tilde{\chi}^0} = \left(
\begin{array} {ccccc}
 M_{BL}   &  0 & 0 & 0 & 0  \\
  0   & M_R & - \frac{1}{2} g_R v_d & \frac{1}{2} g_R v_u & 0 \\
  0   & - \frac{1}{2} g_R v_d & 0 & - \mu & \frac{1}{2} g_2 v_d \\
  0 &  \frac{1}{2} g_R v_u & - \mu & 0 & - \frac{1}{2} g_2 v_u \\
  0 & 0 & \frac{1}{2} g_2 v_d & - \frac{1}{2} g_2 v_u & M_2.
\end{array}
\right)
\end{equation}
Working in the basis where the neutralino mass matrix
is diagonal one finds that the matrix which define
the mixings between the neutrinos and neutralinos in the basis $(\nu_i,\nu^C_j,\tilde{\chi}^0)$
is defined by
\begin{equation}
{\cal M}_{\nu \chi}=\left(
\begin{array} {ccc}
 0 &  M_\nu^D & \Gamma  \\
 (M_\nu^D)^T & 0 & G \\
 \Gamma^T  & G^T & M_{\tilde{\chi}^0} .
\end{array}
\right)
\end{equation}
where
\begin{equation}
\label{Gamma}
\Gamma^{\alpha i}=-\frac{g_{BL}}{2} v_L^\alpha N_{1i}
\ - \ \frac{v_R^\beta}{\sqrt{2}} (Y_\nu^D)^{\alpha \beta} N_{4 i}
\ + \ g_2 v_L^\alpha N_{5 i},
\end{equation}
and
\begin{equation}
\label{G}
G^{\beta i}= \frac{g_{BL}}{2} v_R^\beta N_{1i}
\ - \ \frac{v_L^\alpha}{\sqrt{2}} (Y_\nu^D)^{\beta \alpha} N_{4 i}
\ - \ g_R v_R^\beta N_{2 i}.
\end{equation}
In the above equations, $N$ is the matrix which
diagonalizes the neutralino mass matrix. Now, assuming
that $G, \Gamma << M_{\tilde{\chi}^0}$
integrating out the neutralinos and the right-handed
neutrinos one finds the neutrino mass matrix
\begin{equation}
M_\nu = M_\nu^R \ + \ M_{\nu}^I
\end{equation}
with
\begin{equation}
M_{\nu}^R = - \left( M_\nu^D \ \left( \Gamma G^{-1} \right)^T \ + \ \left( \Gamma G^{-1}\right) (M_\nu^D)^T \right),
\end{equation}
\begin{equation}
M_{\nu}^I = M_\nu^D \ (M_{\nu^C})^{-1} \ (M_\nu^D)^T,
\end{equation}
and
\begin{equation}
M_{\nu^C} = G \ (M_{\tilde{\chi}^0})^{-1} \ G^T.
\end{equation}
In the above equations $M_{\nu}^I$ is the usual
Type I seesaw contribution generated when the
right-handed neutrinos are integrated out, but
the mass matrix for $\nu^C$ is generated
by R-parity violation. Therefore, in this case
neutrino masses are generated through
the double seesaw mechanism. The second contribution,
$M_\nu^R$, is generated by pure R-parity violation.
Therefore, we see that in this theory,
with a mechanism for spontaneous
R-parity violation, it is possible to
generate neutrino masses in a consistent way.
It is important to emphasize that the matrix
$\Gamma$ in Eq.(\ref{Gamma}) and $G$ in Eq.(\ref{G})
can be small and one can have a mini-seesaw mechanism
where the seesaw scale is TeV. In the above equations
$M_\nu^D=Y_\nu^D v_u$, which is, in principle, a free matrix
since the charged lepton masses can be generated
through SUSY loop effects due to the chargino and
neutralino corrections. This is similar to the solution presented
in~\cite{Mohapatra} for the quark sector.
See Ref.~\cite{Barr} for models with similar neutrino mass matrix.
\section{IV. Possible Signals at the LHC}
As it is well known in supersymmetric
scenarios where R-parity is broken the neutralinos
are unstable and new decay channels become available for
the charginos. For a recent analysis of the signals
of R-parity violation see~\cite{Marco}.
In this theory the chargino mass matrix
is given by
\begin{equation}
{\cal M}_{\tilde{\chi}^{\pm}} = \left(
\begin{array} {ccccc}
 M_{R}   &  0 & - \frac{g_R}{\sqrt{2}} v_d   \\
  0   & M_2 & \frac{g_2}{2} v_u \\
 - g_R \frac{v_u}{\sqrt{2}} &  \frac{g_2}{\sqrt{2}} v_d &  \mu
\end{array}
\right)
\end{equation}
when we work in the basis $\tilde{\chi}^+= (\tilde{W}_R^+, \tilde{W}_L^+,\tilde{H}_U^+)$
and $\tilde{\chi}^-= (\tilde{W}_R^-, \tilde{W}_L^-, \tilde{H}_D^-)$.
Now, the matrix that defines the mixing between charged leptons
and charginos reads as

\begin{equation}
{\cal M}_{\pm}=
\left(
\begin{array} {cc}
 M_{\tilde{\chi}^{\pm}}   &  \Gamma^+ \\
 \Gamma^-   &  M_E \\
\end{array}
\right)
\end{equation}
where
\begin{equation}
\Gamma^+_{\alpha i} = \frac{1}{\sqrt{2}} g_2 v_{L\alpha} C_{2i}^+ \ + \
\frac{1}{\sqrt{2}} \left(Y_{\nu}^D \right)_{\alpha \beta} v_R^\beta C_{3i}^+,
\end{equation}
\begin{equation}
\Gamma^-_{\beta i} = -\frac{1}{\sqrt{2}} g_R v_{R \beta} C_{1i}^-  \ + \
\frac{1}{\sqrt{2}} \left(Y_{\nu}^D \right)_{\alpha \beta} v_L^\alpha C_{3i}^-.
\end{equation}
Here, $C^{\pm}$ are the matrices which diagonalize the chargino mass matrix.
The generic predictions coming from R-parity scenarios are
the decays of neutralinos and the new decays for the charginos.
In our case we have three charginos, $\tilde{\chi}^{\pm}$, which will have the
following decay channels: $\tilde{\chi}^{\pm}_i \ \to \ e^{\pm}_j Z, \ \nu W^{\pm}$
through the coupling $\Gamma^{\pm}$, and the neutralinos decays
$\tilde{\chi}^0_i \to \nu Z, e^{\pm}_j W^{\mp}$ through the coupling
$\Gamma$ and $G$, respectively. Therefore, once we take into account the neutrino
mass constraints one can predict these decays~\cite{Spinner}.

It is important to mention that once the charginos are integrated out
one can generate mass for one charged lepton. In this case:
\begin{equation}
(M_E)_{\alpha \beta} = \Gamma_{\alpha i}^+ \ M_{\tilde{\chi}^+_i}^{-1} \ \Gamma_{\beta i}^-.
\end{equation}
And, neglecting the terms proportional to $Y_\nu^D$:
\begin{equation}
(M_E)_{\alpha \beta} \approx - \frac{g_2 g_R}{2} v_{L \alpha} v_{R \beta} \frac{C_{2i}^+ C_{1i}^-}{M_{\tilde{\chi}_i}}.
\end{equation}
Therefore, one could generate one of the charged lepton masses once the charginos are integrated out.
There are some new novel decays in this theory. For example the decays $\tilde{\nu} \to e^-_i e^+_j$,
$Z^{'} \ \to \ e_i^{\pm} \tilde{\chi}_j^{\mp}$ and $W_R^{\pm} \ \to \ e_i^{\pm} \tilde{\chi}^0_j$
which could help test this theory. Before finishing this section, we would like
to emphasize that in this case the R-parity violating decays of the neutralinos
and charginos are not highly suppressed by neutrino masses since they are
proportional to the couplings $\Gamma$, $G$
and $\Gamma^{\pm}$. This is an important difference between the usual
R-parity violating scenarios and this one. We will study these issues in great detail
in a future publication~\cite{Spinner}.
\section{Summary and Outlook}
We have investigated the connection between R-parity and the possibility
of finding the simplest supersymmetric left-right symmetric theory.
We found a simple theory where R-parity is spontaneously
broken and neutrino masses are generated through Type I seesaw and R-parity violation.
In this theory R-parity and the $SU(2)_R$ symmetry are broken by the vacuum expectation value of
the sneutrinos, which are related to the SUSY breaking scale. The Higgs sector of the theory is
quite simple since could be composed of the MSSM Higgses or only two bidoublets.
We have discussed the spectrum of the theory, and the possible tests
at the Large Hadron Collider though the decays of neutralinos, charginos, $Z^{'}$ and $W_R^{\pm}$.
Furthermore, neutralinos and charginos decays are not highly suppressed by neutrino
masses because of the double seesaw mechanism. The phenomenological and cosmological
aspects of this theory will be investigated in detail in a future publication.

{\textit{Acknowledgments}}.
We would like to thank M. Drees, R.N. Mohapatra and G. Senjanovi\'c
for useful discussions and careful reading of the manuscript.
We thank V. Barger for discussions. The work of P. F. P. was
supported in part by the U.S. Department of Energy contract
No. DE-FG02-08ER41531 and in part by the Wisconsin Alumni
Research Foundation. S.S. is supported in part by the U.S.
Department of Energy under grant No. DE-FG02-95ER40896,
and by the Wisconsin Alumni Research Foundation.
\appendix


\end{document}